\documentclass[prl,aps,showpacs,floatfix,nofootinbib,twocolumn,letterpaper]{revtex4}
\usepackage{epsfig}
\usepackage{graphicx}
\usepackage{amssymb}
\usepackage{amsmath}
\usepackage{amsfonts}

\newcommand{\be}{\begin{equation}}
\newcommand{\ee}{\end{equation}}
\newcommand{\ba}{\begin{eqnarray}}
\newcommand{\ea}{\end{eqnarray}}

\newcommand{\Tr}{{\rm Tr}}

\begin{document}
\title{Direct microscopic calculations of nuclear level densities in the shell model Monte Carlo approach}
\author{Y. Alhassid$^{1}$, M. Bonett-Matiz$^{1}$, S. Liu$^{1}$, and H. Nakada$^{2}$}
\affiliation{$^{1}$Center for Theoretical Physics, Sloane Physics
Laboratory, Yale University, New Haven, CT 06520\\
$^{2}$ Department of Physics, Graduate School of Science, Chiba University, Inage, Chiba 263-8522, Japan}
\date{\today}
\begin{abstract}
Nuclear level densities are crucial for estimating statistical nuclear reaction rates. The shell model Monte Carlo method is a powerful approach for microscopic calculation of state densities in very large model spaces. However, these state densities include the spin degeneracy of each energy level, whereas experiments often measure level densities in which each level is counted just once.  To enable the direct comparison of theory with experiments, we introduce a method to calculate directly the level density in the shell model Monte Carlo approach. The method employs a projection on the minimal absolute value of the magnetic quantum number. We apply the method to nuclei in the iron region as well as the strongly deformed rare-earth nucleus $^{162}$Dy. We find very good agreement with experimental data including level counting at low energies, charged particle spectra and Oslo method at intermediate energies, neutron and proton resonance data, and Ericson's fluctuation analysis at higher excitation energies.
\end{abstract}
\pacs{21.10.Ma, 21.60.Cs, 21.60.Ka, 21.60.De}
\maketitle
\emph{Introduction.} The level density is among the most important statistical properties of the atomic nucleus. It appears explicitly in Fermi's golden rule for transition rates and in the Hauser-Feshbach theory~\cite{Hauser1952} of statistical nuclear reactions.  Yet its microscopic calculation presents a major theoretical challenge. In particular, correlations have important effects on nuclear level densities but are difficult to include quantitatively beyond the mean-field approximation. The configuration-interaction (CI) shell model is a suitable framework to include both shell effects and correlations. However, the dimension of the required model space increases combinatorially with the number of single-particle states and/or the number of nucleons, and conventional shell model calculations become intractable in medium-mass and heavy nuclei. This difficulty has been overcome using the shell model Monte Carlo (SMMC) approach~\cite{Lang1993,Alhassid1994,Koonin1998,Alhassid2001}. The SMMC has proved to be a powerful method to calculate microscopically nuclear state densities~\cite{Nakada1997,Ormand1997,Alhassid1999,Langanke1998,Alhassid2007,Ozen2007}.

The SMMC method is based on a thermodynamic approach, in which observables such as the thermal energy are calculated by tracing over the complete many-particle Hilbert space. Thus, the SMMC {\em state} density takes into account the magnetic degeneracy of the nuclear levels so that each level of spin $J$ is counted $2J+1$ times.

However, experiments often measure the {\em level} density, in which each level is counted exactly once, irrespective of its spin degeneracy~\cite{Dilg1973,Iljinov1992}. To make direct comparison of theory with experiments, it is thus necessary to be able to calculate the level density within the SMMC approach. A spin-projection method, introduced in Ref.~\onlinecite{Alhassid2007}, can be used to calculate the level density $\rho_J(E_x)$ for a given spin $J$ at each excitation energy $E_x$. While the state density is given by  $\rho(E_x)= \sum_J (2J+1) \rho_J(E_x)$, the total level density is $\tilde\rho(E_x)= \sum_J \rho_J(E_x)$. However, this latter formula is not useful for practical calculations because the statistical errors of $\rho_J(E_x)$ increase with $J$, and the resulting statistical errors in $\tilde\rho(E_x)$ are too large.

Here we introduce a simple method to calculate directly and accurately the level density in SMMC. We present level density calculations of medium-mass nuclei in the iron region and of the well-deformed nucleus $^{162}$Dy. We find good agreement with a variety of experimental data including level counting at low energies, charged particle spectra and Oslo method at intermediate energies, neutron and proton resonance data, and Ericson's fluctuation analysis at higher excitation energies. We note that our method can be applied more generally to many-particle systems with good total angular momentum.

\emph{Level density in SMMC}.  We make the observation that for any nuclear level with spin $J$ and magnetic quantum number degeneracy of $2J+1$, the state with the lowest possible non-negative spin projection $M$ appears exactly once. Denoting by $\rho_M$ the level density for a given value of the spin projection $M$, the total level density for even-even and odd-odd nuclei (whose spin is integer) is given by $\tilde\rho=\rho_{M=0}$, while for odd-even nuclei (whose spin is half-integer), the total level density is $\tilde\rho=\rho_{M=1/2}$.

The $M$-projected level density can be calculated as in Ref.~\onlinecite{Alhassid2007}. For a nucleus described by a shell model Hamiltonian $H$ and at inverse temperature $\beta=1/T$, the SMMC method is based on the Hubbard-Stratonovich (HS) transformation~\cite{HS-trans} $e^{-\beta  H} = \int D[\sigma] G_\sigma U_\sigma$, where $G_\sigma$ is a Gaussian weight and $U_\sigma$ is a one-body propagator describing non-interacting nucleons in time-dependent auxiliary fields $\sigma$. For a quantity $X$ that depends on the auxiliary fields $\sigma$, we define
\be
\overline{{X}_{\sigma}} \equiv \frac {\int D[\sigma ] W(\sigma) X_{\sigma} \Phi_{\sigma}} { \int D[\sigma] W(\sigma) \Phi_{\sigma}} \;,
\ee
where $W(\sigma) = G_\sigma |\Tr\, U_{\sigma}|$ is the weight used in the Monte Carlo sampling and $\Phi_{\sigma}=\Tr\, U_{\sigma}/|\Tr\, U_{\sigma}|$ is the Monte Carlo sign function. Here and in the following, the traces are evaluated in the canonical ensemble for fixed number of protons and neutrons, which in turn can be calculated from grand-canonical traces by particle-number projection.

The  $M$-projected thermal energy $E_M(\beta) = \langle H \rangle_M$ is calculated using
\be \label{M-observ}
\langle H \rangle_M  \equiv {{\rm Tr}_M \left( H e^{-\beta H} \right) \over {\rm Tr}_M e^{-\beta H }}
={\overline{\big[{{\rm Tr}_M
(H U_\sigma)  \over {\rm Tr} U_\sigma}\big]}
\over \overline{\big[{{\rm Tr}_M U_\sigma  \over
{\rm Tr}\, U_\sigma}\big]} } \;.
\ee

The trace  ${\rm Tr}_M X$ at fixed spin component $M$ can be calculated by a discrete Fourier transform
\begin{equation}
  \label{M-project}
  {\rm Tr}_M X = {1 \over 2J_s + 1} \sum\limits_{k =-J_s}^{J_s}
  e^{-i \varphi_k M} {\rm Tr} \left( e^{i\varphi_k \hat J_z} X \right) \;,
\end{equation}
where $\varphi_k$ ($k=-J_s,\ldots, J_s$) are quadrature points $\varphi_k=\pi {k \over J_s+1/2}$ and $J_s$ is the maximal spin in the many-particle shell model space.

The $M$-projected canonical partition function $Z_M(\beta)$ is calculated by integrating the thermodynamic relation $-d\ln Z_M /d\beta = E_M(\beta)$, taking $Z_M(\beta=0)$ to be the total number of levels with the magnetic quantum number $M$. For the lowest non-negative value of $M$, $Z_M(\beta=0)$ is the total number of levels without counting their magnetic degeneracy. The $M$-projected level density $\rho_M(E_x)$ is then calculated in the saddle-point approximation
\be\label{eq:rho_M}
\rho_M  \approx {1 \over \sqrt{2\pi T^2 C_M}} e^{S_M} \;,
\ee
where $S_M$ and $C_M$ are, respectively, the $M$-projected canonical entropy and heat capacity
\be\label{eq:S-C}
 S_M =\ln Z_M + \beta E_M \;;\;\;\; C_M=\frac{dE_M}{dT}=-\beta^2 {dE_M \over d\beta} \;.
\ee
 In the calculation of $C_M$ we implemented the method of Ref.~\onlinecite{Liu2001}, in which the numerical derivative is carried out inside the HS path integral. This enable us to take into account correlated errors, thus reducing significantly the statistical errors in the heat capacity compared to a direct numerical derivative of the thermal energy.  Equation~(\ref{eq:rho_M}) is analogous to the formula used for the state density~\cite{Nakada1997} in which the corresponding quantities do not include $M$ projection.

The projection on the spin component $M$ usually introduces a sign problem that leads to large fluctuations of observables at low temperatures (even for a good-sign interaction).  However, for even-even nuclei ${\rm Tr} \left( e^{i\varphi_k \hat J_z} U_\sigma \right)$ is almost always positive (for a good-sign interaction), and using Eq.~(\ref{M-project}) with $M=0$ and $X=U_\sigma$ we have ${\rm Tr}_{M=0} U_\sigma > 0$. Thus the level density of even-even nuclei can be calculated accurately down to low excitation energies without a sign problem.

\emph{Medium-mass nuclei}. We demonstrate the SMMC calculation of level densities for medium-mass nuclei in the iron region using the CI shell model Hamiltonian of Ref.~\onlinecite{Nakada1997} in the complete $pfg_{9/2}$ shell.

\begin{figure}[t]
\includegraphics[width=0.9\columnwidth]{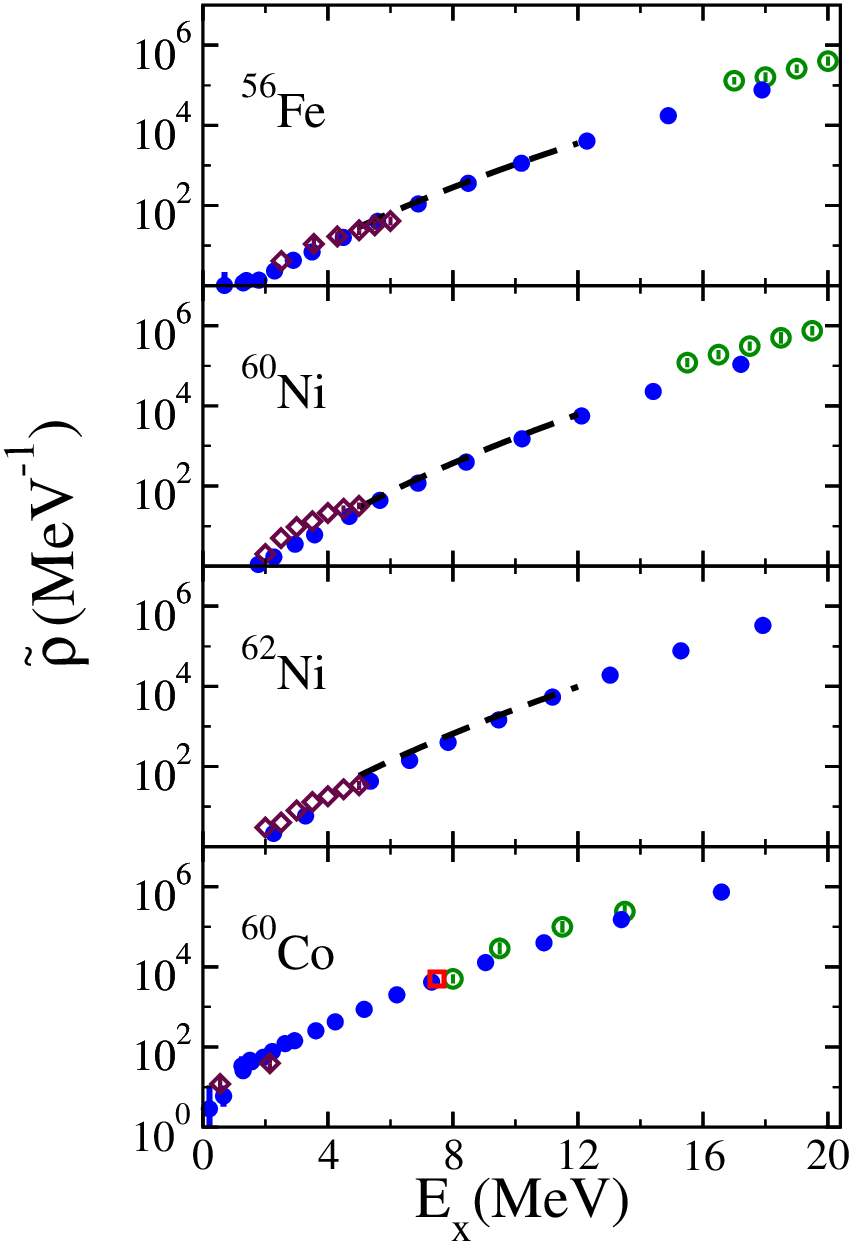}
\caption{Level densities versus excitation energy $E_x$ for $^{56}$Fe, $^{60}$Ni, $^{62}$Ni and $^{60}$Co.  SMMC level densities $\tilde\rho(E_x)=\rho_{M=0}(E_x)$ (solid circles) are compared with various experimental data sets~\cite{Iljinov1992}: level counting at low excitation energies (open diamonds), charged particle spectra~\cite{Lu1972} at intermediate energies (dashed lines),  and Ericson's fluctuation analysis~\cite{Huizenga1969} at higher energies (open circles). For $^{60}$Co there is also the proton resonance data (open square)~\cite{Lindstrom1971,Browne1970}.}
\label{fpg-level}
\end{figure}

In Fig.~\ref{fpg-level} we compare SMMC level density calculations (solid circles with error bars) for $^{56}$Fe, $^{60}$Ni, $^{62}$Ni and $^{60}$Co with various experimental data compiled in Ref.~\onlinecite{Iljinov1992}: (i) level counting at low excitation energies (open diamonds), (ii) charged particle reactions such as $(\alpha,\alpha')$, $(p,p')$, $(p,\alpha)$ and $(\alpha,p)$ at intermediate excitation energies (dashed lines)~\cite{Lu1972}, and (iii) Ericson's fluctuation analysis at higher excitation energies (open circles)~\cite{Huizenga1969}. For $^{60}$Co there is also high-resolution proton resonance data at around $8$ MeV (open square)~\cite{Lindstrom1971,Browne1970}. Overall, we find good agreement between the SMMC calculations and the experimental data.

\emph{Spin-cutoff parameter}.  In the spin-cutoff model, the spin distribution $\rho_J(E_x)$ is given by
\be
  \label{spin-cutoff}
  \rho_J(E_x) = \rho(E_x)
  {(2J+1) \over 2\sqrt{2 \pi} \sigma_c^3} e^{-{J(J+1) \over 2
      \sigma_c^2}}\;,
\ee
where $\rho(E_x)$ is the total state density and $\sigma_c=\sigma_c(E_x)$ is an energy-dependent spin-cutoff parameter. The distribution (\ref{spin-cutoff}) is normalized such that $\sum_J (2J+1) \rho_J(E_x) \approx \rho(E_x)$. Equation (\ref{spin-cutoff}) can be derived in the random coupling model of individual spins~\cite{Ericson1960}. In this model, the level density $\tilde \rho(E_x)$ can be calculated to be
\be
\tilde \rho(E_x) = \sum_J \rho_J(E_x) \approx {1 \over \sqrt{2\pi} \sigma_c} \rho(E_x) \;,
\ee
where the sum over spin is calculated by converting it to an integral.
An effective spin-cutoff parameter can then be estimated from the ratio of the total state density to the total level density, i.e., $\sigma_c(E_x) = (2\pi)^{-1/2}\rho(E_x)/\tilde\rho(E_x)$. The spin-cutoff parameter can be converted to a thermodynamic moment of inertia $I$ using $\sigma_c^2 = I T/\hbar^2$.

\begin{figure}[t]
\includegraphics[width=0.85\columnwidth]{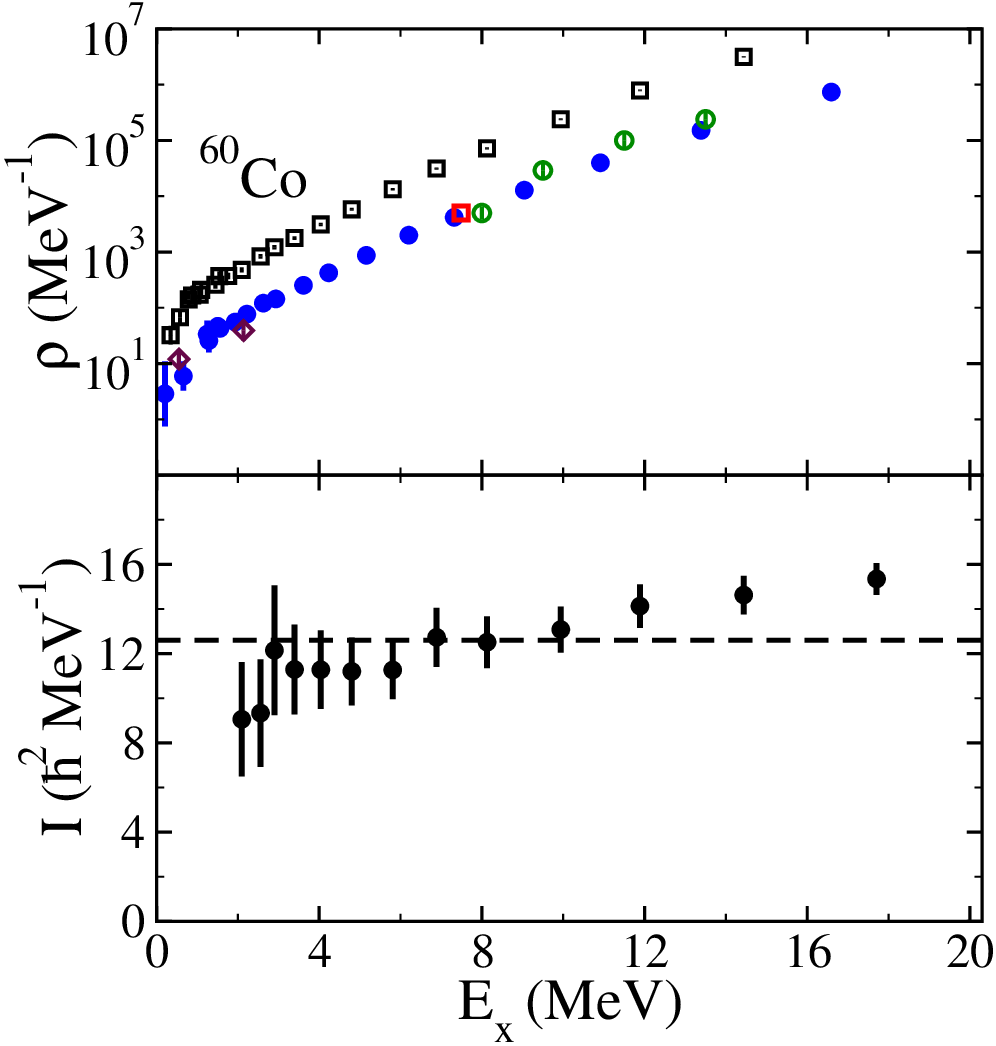}
\caption{Top: The SMMC state density (open squares) and level density (solid circles) versus excitation energy $E_x$ for $^{60}$Co. The experimental level density data follows the same convention of Fig.~\ref{fpg-level}. Bottom: thermodynamic moment of inertia for $^{60}$Co extracted from the ratio of the state density to the level density (solid circles). The dashed line is the rigid-body moment of inertia.}
\label{Co60-inertia}
\end{figure}

\begin{figure}[]
\includegraphics[width=0.85\columnwidth]{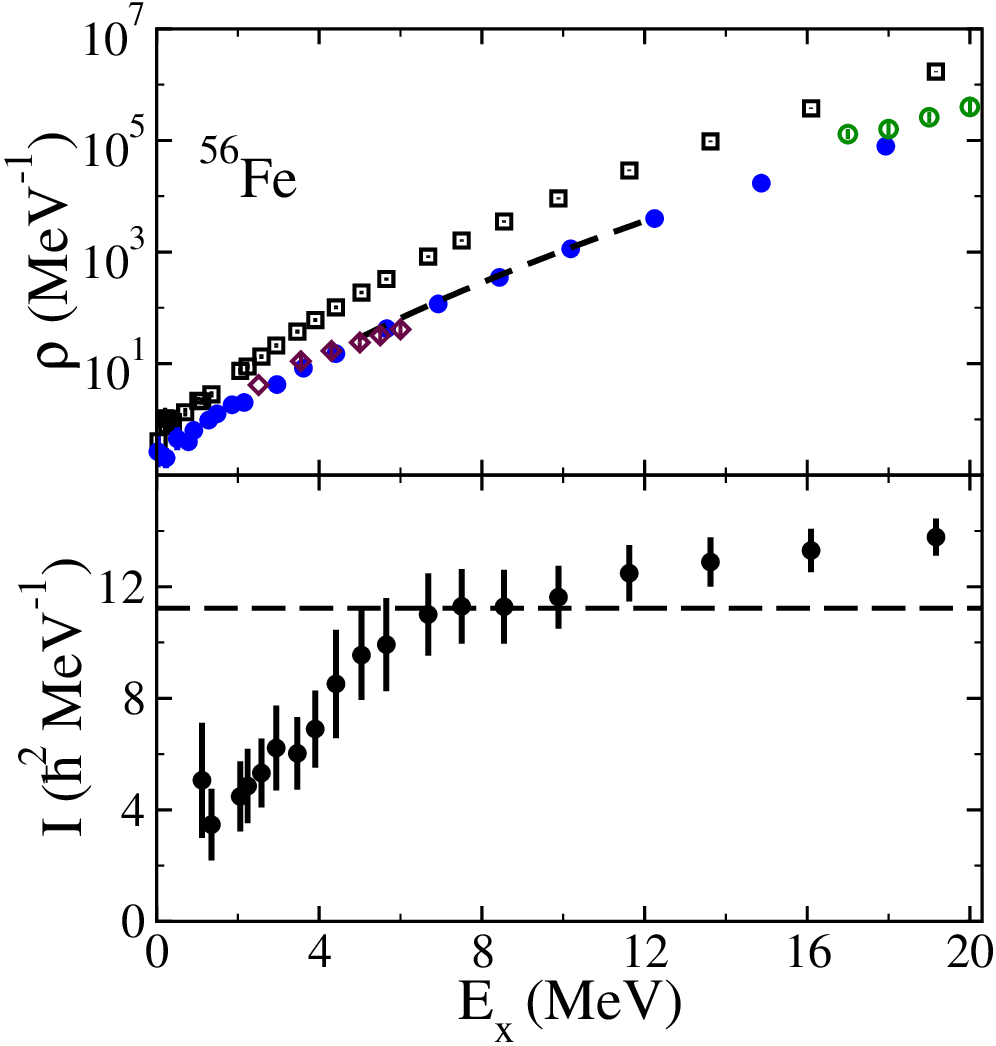}
\caption{As in Fig.~\ref{Co60-inertia} but for $^{56}$Fe.}
\label{Fe56-inertia}
\end{figure}

We have extracted such moment of inertia $I$ from the calculated SMMC state and level densities of $^{56}$Fe and $^{60}$Co. In Figs.~\ref{Co60-inertia} and \ref{Fe56-inertia} we show the corresponding state densities (open squares) and level densities (solid circles) and the corresponding moment of inertia $I$ (bottom panels) versus excitation energy $E_x$. For the odd-odd nucleus $^{60}$Co the moment of inertia depends only weakly on excitation energy. However, for the even-even nucleus $^{56}$Fe we observe a suppression of the moment of inertia at low excitation energies. This reflects the reduction in the state-to-level density ratio that originates in pairing correlations, and is consistent with the results found in Ref.~\onlinecite{Alhassid2007} in which the moment of inertia was extracted from the spin distribution.


\emph{Rare-earth nucleus $^{162}$Dy}. In Refs.~\onlinecite{Alhassid2008} and \onlinecite{Ozen2012}  we extended the SMMC approach to heavy nuclei in the rare-earth region using the 50-82 major shell plus the $1f_{7/2}$ orbital for protons, and the 82-126 major shell plus $0h_{11/2}$ and $1g_{9/2}$ orbitals for neutrons. We described successfully the rotational character of the strongly deformed nucleus  $^{162}$Dy~\cite{Alhassid2008} as well as the crossover from vibrational to rotational collectivity in families of samarium and neodymium isotopes~\cite{Ozen2012}.

\begin{figure}[t]
\includegraphics[width=\columnwidth]{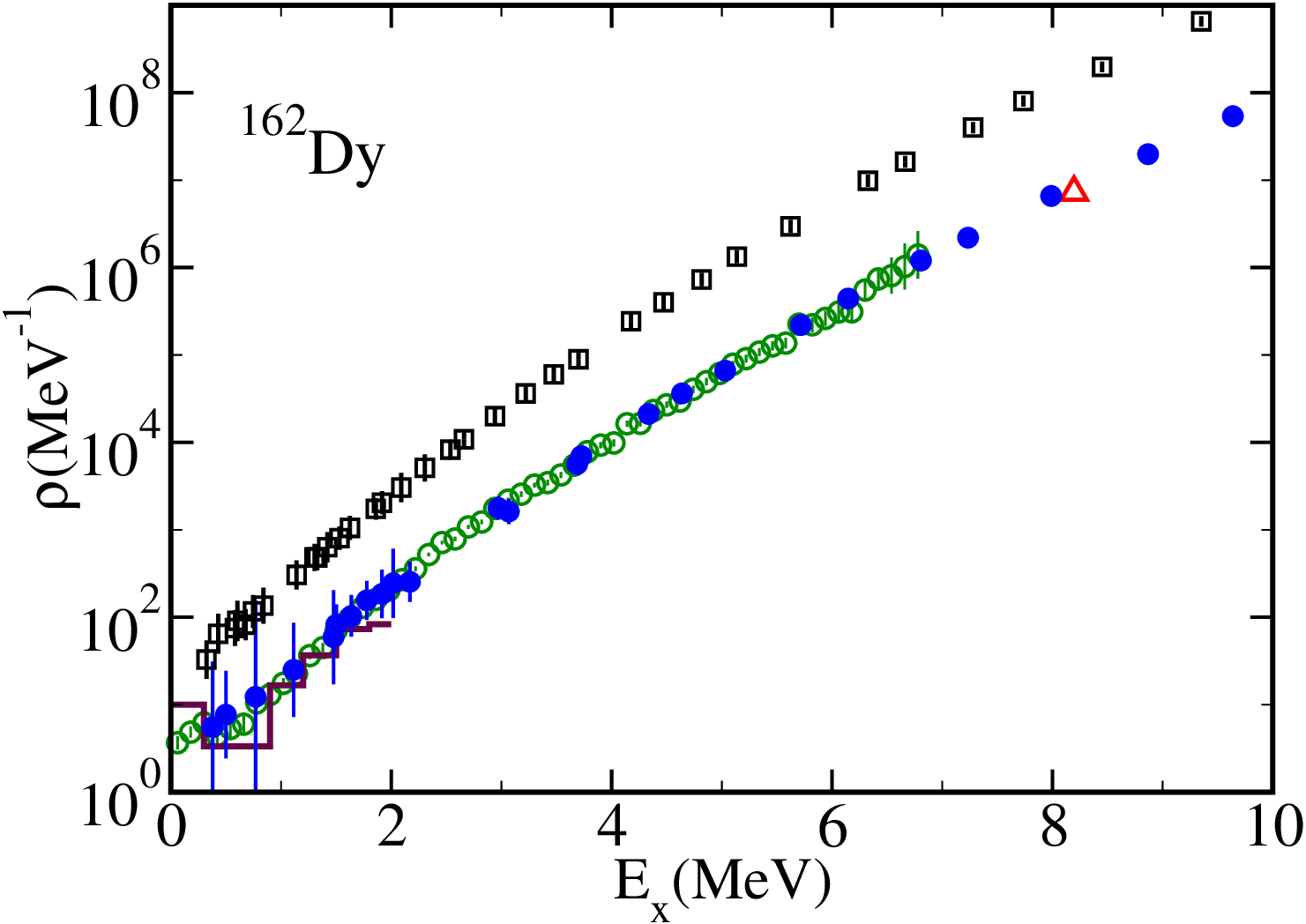}
\caption{Level density and state density in $^{162}$Dy. The SMMC level density (solid circles) is compared with the state density (open squares).  Also shown are experimental data sets for the level density: level counting at low excitation energies (histograms)~\cite{Isotope,Aprahamian2006}, Oslo data at intermediate energies (open circles)~\cite{Oslo2000,Guttormsen2003}, and the neutron resonance data (triangle)~\cite{IAEA1998}.}
\label{Dy162-level}
\end{figure}

Fig.~\ref{Dy162-level} shows the SMMC level density $\tilde\rho(E_x) = \rho_{M=0}(E_x)$ (solid circles) and SMMC state density $\rho(E_x)$ (open squares) for $^{162}$Dy. The SMMC level density compares well with various experimental data sets: (i) level counting (solid histograms)~\cite{Isotope,Aprahamian2006}, (ii) renormalized Oslo data (open circles)~\cite{Oslo2000,Guttormsen2003} and (iii) neutron resonance data (triangle)~\cite{IAEA1998}. We find very good agreement between the various data sets and the SMMC level density.

\emph{Conclusion}.  In conclusion, we have used a spin-component projection method to calculate directly and accurately the SMMC nuclear level density $\tilde\rho(E_x)$ as the projected density $\rho_{M=0}(E_x)$ for even-even and odd-odd nuclei. The method is easily extended to odd-even nuclei by using $\tilde\rho(E_x)=\rho_{M=1/2}(E_x)$. This method allows us to make direct comparison with experimental data. We find very good agreement between the SMMC level density and the experimental data for nuclei in the iron region and for the rare-earth nucleus $^{162}$Dy.

\emph{Acknowledgements.} This work was supported in part by the U.S. Department of Energy Grant No. DE-FG02-91ER40608, and by the JSPS Grant-in-Aid for Scientific Research (C) No.~25400245. Computational cycles were provided by the he NERSC high performance computing facility at LBL and by the High Performance Computing Center at Yale University.

\end{document}